# Solvation Effects on Hole Mobility in the Poly G/Poly C Duplex

V. D. Lakhno and N. S. Fialko

*Institute of Mathematical Problems of Biology, Russian Academy of Sciences,
Pushchino, Moscow oblast, Russia*

*e-mail: lak@impb.psn.ru; fialka@impb.psn.ru*

**Abstract**—Theoretical calculations of solvation contribution to hole energy in a polynucleotide chain give very low hole mobility values at zero temperature, $\mu < 10^{-3}$ cm$^2$/(V s). We calculated hole mobility at physiological temperature for the Poly G/Poly C DNA duplex, which gave substantially larger mobility values. Mobility over the temperature range 20–400 K was calculated. Taking stacking interaction into account substantially increased hole mobility.

## INTRODUCTION

Studies of charge transfer in DNA are of importance for both biophysics (the propagation of foreign charges along DNA is part of the mechanism of such important biochemical processes as replication, transcription, decomposition and reparation of DNA, and movement of radicals over DNA molecules, which play an important role in mutagenesis and cancerogenesis processes) and nanobioelectronics, a new rapidly developing discipline integrating achievements in nanoelectronics and molecular biology. The main directions in nanobioelectronics include the development of biosensors, DNA-based nanoelectronic circuits, biochips, nanomotors, nanotransporters, etc. For this reason, studies of the conducting properties of DNA molecules are of great interest [1, 2].

Mathematical simulation of charge transfer in biological systems is related to the use of discrete models, in which charge transfer paths in macromolecules are considered [3–6]. Such models take into account various possible particle interactions with a chain, but are constructed on general principles. A molecular chain is modeled by a discrete chain consisting of sites (sites are molecules or groups of strongly bound atoms; sites interact with each other by comparatively weak forces). Excess charge (electron or hole) is introduced into the chain.

The state with a hole localized on one site is not an energetically optimum configuration. Indeed, localization results in a substantial kinetic energy loss. This loss decreases when the hole is delocalized over several chain sites. In reality, this effect is similar to the mechanism of polaron formation in ionic crystals, when delocalization over several sites contributes to a gain in the total polaron energy. Charge propagation influences movement of sites. Conversely, site displacement changes the probability of the presence of charge on it. Charge movement is described by the Schrödinger equation, and site oscillations, by classical equations of motion.

Simulation of a chain of sites is performed by considering two types of interactions in DNA: between complementary bases in one plane on neighboring strands (interaction through H-bond formation) and between bases lying on one strand one above another in a "stack," stacking interaction caused by several reasons such as dispersion London forces, hydrophobic effects, and overlap of aromatic ring π orbitals present in bases [7]. To take stacking into account (the Peyrard–Bishop model [8]), the "total" potential of these forces not divided into components is considered. The term describing stacking interaction in model equations has the form similar to that of the dispersion term in equations for crystals.

The importance of the inclusion of dispersion (stacking interaction) in charge transfer in a nucleotide chain was noted in a large number of works on charge transfer in DNA, but, so far as we know, no quantitative estimate of the role played by intersite interaction in hole mobility at room temperature was obtained.

In molecular crystals, the ξ dispersion value is usually small, $\xi \ll 1$. This is not so with DNA. As distinct from molecular crystals, a very important role is played in DNA by stacking interaction, which determines dispersion in a classic chain. The stacking interaction value in DNA was found to be $K_D = 0.04$ eV/Å$^2$ [9]. For intramolecular vibrations (interaction constant with stretching and contraction H-bond deformations), $K \approx 0.06$ eV/Å$^2$ was obtained. The ratio between the parameter values corresponding to the poly(G)/poly(C) duplex [10] calculated theoretically with and without taking dispersion (stacking) into account was $\mu/\mu_0 \approx 2.79$ ($\mu_0$ is the hole mobility in a polyguanine chain without dispersion, and $\mu$ is the mobility with dispersion taken into account). This is evidence of sensitivity of the conduction properties of DNA to the stacking interaction value.

Solvation effects play an important role in transfer processes. Various authors give different theoretical mobility estimates. Charge solvation decreases the rate of transfer by 7–8 orders of magnitude compared with "dry" DNA [11]. Solvated hole mobility at $T = 0$ is very low, approximately $2 \times 10^{-4}$ cm$^2$/(V s) [12]. The results obtained in [13] were as follows: the interaction energy between a hole and polarized water molecules ~0.5 eV, a hole is "spread" over 3–5 sites, and hole mobility does not exceed $10^{-2}$–$10^{-3}$ cm$^2$/(V s). In [14], a higher interaction characteristic (solvation energy) was obtained, 1.02 eV; a charge in the DNA chain was localized on one site with a probability higher than 0.9.

In this work, we numerically studied the influence of solvation effects and dispersion in a classic chain on hole mobility at a temperature close to a physiological temperature. Model parameters were taken for a uniform guanine chain. The simplification made was the assumption that parameter values did not change as the temperature varied.

## MODEL

The model was based on the Holstein Hamiltonian for a discrete chain of sites [10, 15],

$$\hat{H} = \sum_{m,n} \nu_{mn} |m\rangle\langle n| + \tilde{\chi} \sum_n \tilde{u}_n |n\rangle\langle n| + \sum_n \Phi_n |n\rangle\langle n| \quad (1)$$
$$+ \sum_n \frac{M \dot{\tilde{u}}_n^2}{2} + \sum_n \frac{K \tilde{u}_n^2}{2} + \sum_n \frac{K_D(\tilde{u}_n - \tilde{u}_{n-1})^2}{2}.$$

With DNA, a nucleotide pair is considered a site. Here, $\nu_{mn}$ ($m \neq n$) denotes matrix elements of electron transfer between the $m$th and $n$th sites, $\nu_{nn}$ is the charge energy on the $n$th site, $\tilde{\chi}$ is the interaction constant between an electron and $n$th site displacements $\tilde{u}_n$ from its equilibrium position, $\Phi_n$ is the charge solvation energy on the $n$th site, $M$ is the effective mass of the site, $K$ is the elastic constant, and $K_D$ is the constant determining the contribution of dispersion to chain energy.

If the $\Psi$ wave function is selected in the form $\Psi = \sum_{n=1}^N b_n |n\rangle$, where $b_n$ is the probability amplitude of the occurrence of a charge on the $n$th site, and if $\Phi_n = \tilde{\zeta} q_n$ [6], where $q_n = |b_n|^2$ is the charge distribution density on the $n$th site and $\tilde{\zeta}$ is the effective solvation coefficient, Eq. (1) gives the total energy of the system $\langle\Psi|\hat{H}|\Psi\rangle$ in the form

$$\langle\Psi|\hat{H}|\Psi\rangle = \sum_{m,n} \nu_{mn} b_m b_n^* + \sum_i \frac{M \dot{\tilde{u}}_i^2}{2} + \sum_i \frac{K \tilde{u}_i^2}{2}$$
$$+ \sum_i \frac{K_D(\tilde{u}_i - \tilde{u}_{i-1})^2}{2} + \sum_n \tilde{\chi} \tilde{u}_n b_n b_n^* + \sum_n \frac{\tilde{\zeta}}{2}(b_n b_n^*)^2. \quad (2)$$

We use the nearest neighbor approximation; that is, $\nu_{mn} = 0$ if $m \neq n \pm 1$. An important role is played in DNA by nonlinear stacking interaction $\Xi(\tilde{u}_n - \tilde{u}_{n-1})$ [8, 9]. At small $(\tilde{u}_n - \tilde{u}_{n-1})$ difference values, this interaction is given in the first approximation by the quadratic dependence $\Xi(\tilde{u}_n - \tilde{u}_{n-1}) = (K_D/2)(\tilde{u}_n - \tilde{u}_{n-1})^2$.

Equations of motion obtained from Hamiltonian (2) and written in the dimensionless form are

$$i\dot{b}_n = \eta_{n,n-1} b_{n-1} + \eta_{n,n} b_n + \eta_{n,n+1} b_{n+1} \quad (3)$$
$$+ \chi u_n b_n + \varsigma b_n |b_n|^2,$$

$$\ddot{u}_n = -(\omega^2 + 2\xi) u_n + \xi(u_{n+1} + u_{n-1}) \quad (4)$$
$$- \chi |b_n|^2 - \gamma \dot{u}_n + A_n(t).$$

Dimensionless values are related to dimensional parameters as follows. Let us select characteristic time $\tau$, $\tilde{t} = \tau t$, and characteristic scale of vibrations $U^*$, $\tilde{u}_n = U^* u_n$. This gives $\eta_{nm} = \nu_{nm} \tau/\hbar$ for matrix elements, $\zeta = \tilde{\zeta} \tau/\hbar$ for the solvation coefficient, $\omega = (\tau^2 K/M)^{1/2}$ for site vibrational frequencies, and $\xi = (\tau^2 K_D/M)^{1/2}$ for the dispersion coefficient. Making the $\tilde{\chi}$ coupling constant between the quantum and classical system parts dimensionless requires that the coefficients of $u_n b_n$ in (3) and $|b_n|^2$ in (4) be equal. We then have $U^* = (\tau \hbar/M)^{1/2}$ and $\chi = \tilde{\chi}(\tau^3/\hbar M)^{1/2}$.

We added a term with friction $\gamma = \tilde{\gamma} \tau/M$ ($\tilde{\gamma}$ is the friction coefficient) and random force $A_n(t)$ with the properties $\langle A_n(t)\rangle = 0$, $\langle A_n(t) A_k(t')\rangle = 2k_B T(\tilde{\gamma} \tau^3/M^2 U^{*2}) \delta_{kn} \delta(t - t')$. These terms simulate the environment with the required temperature $T$ (the Langevin equation).

## A METHOD FOR MOBILITY CALCULATIONS

The numerical integration of system (3), (4) with given initial conditions describes the dynamics of a charge and the trajectories of properties at given temperature in a separate realization. We calculate the mobility of a hole in a chain using the method described in [16], that is, calculate a set of realizations (numerically integrate the system with different initial data corresponding to distributions of displacements and velocities of classical chain sites at given temperature $T$; the charge at the initial time is considered

Mobility μ, cm$^2$(V s), at $T_0 = 300$ K

| ξ | ζ | μ | ξ | ζ | μ |
|---|---|---|---|---|---|
| | 1a | | | 1b | |
| 0 | 0 | 2.87 | $6.4 \times 10^{-5}$ | 0 | 12.6 |
| | 2a | | | 2b | |
| 0 | −15.5 | 0.052 | $6.4 \times 10^{-5}$ | −15.5 | 0.25 |

localized in the center of a chain on site number 0, $b_0(t=0) = 1$. Next, we find the mean square displacement averaged over realizations,

$$\langle X^2(t) \rangle = \left\langle \sum_{n=-[N/2]}^{[N/2]} |b_n(t)|^2 n^2 \right\rangle$$

and use it to approximate diffusion coefficient $D$, $\langle X^2(t) \rangle = 2Dt$, and charge mobility $\mu = eD/k_B T$ at temperature $T$ ($e$ is the charge of the electron).

We considered uniform synthetic polyguanine fragments. Model parameters corresponded to the guanine chain [17–20]: $\tau = 10^{-14}$ s, $M = 10^{-21}$ g, $\eta = 1.276$ ($\nu = 0.084$ eV), $\omega = 0.01$ (picosecond site vibrational frequency $\tilde{\omega} = 10^{12}$ s$^{-1}$ corresponded to rigidity $K \sim 0.062$ eV/Å$^2$), $\chi = 0.02$ ($\tilde{\chi} = 0.13$ eV/Å), and $\gamma = 0.006$ ($\tilde{\gamma} = 6 \times 10^{11}$ s$^{-1}$). The initial data for classical sites corresponded to equilibrium distribution at given temperature. The charge at the initial time was localized on one site in the center of the chain.

Calculations were performed for four variants. First, for a chain without taking solvation into account, a model of dry DNA without solvation effects:

(1a) A model without dispersion in a chain of sites and without solvation (ξ = 0 and ζ = 0).

(1b) A model with dispersion interactions (ξ = $6.4 \times 10^{-5}$ ($K_D = 0.04$ eV/Å$^2$), ζ = 0).

Secondly, for a chain with solvation simulating DNA in solution:

(2a) A model with solvation but without dispersion (ξ = 0, ζ = −15.5 ($\tilde{\zeta} = -1.04$ eV)).

(2b) A model with dispersion and solvation (ξ = $6.4 \times 10^{-5}$ and ζ = −15.5).

Calculations were performed over the temperature range 20–400 K, in steps of 50 K from 100 to 400 K and in steps of 20 K at low temperatures, from 20 to 100 K. The mean was calculated for 500 realizations.

## RESULTS OF MOBILITY CALCULATIONS

The table contains calculated mobilities for all the variants at a "biologically significant" temperature (300 K).

Mobility calculations up to 400 K showed that, although mobility values obtained using various calculation variants can be substantially different (see variants 1b and 2a, mobility values differ by two orders of magnitude), the temperature dependences of mobility at lower temperatures are similar in all cases (Fig. 1); that is, the temperature dependence of mobility is close to the $(\mu/\mu_0) = (T/T_0)^{-2.2}$ dependence [21].

The dynamics of transfer is strongly different for systems with and without solvation. In a chain without solvation (1a and 1b), charge fairly quickly spreads from the initial state (1 on one site in the center of the chain and 0 on all the other sites). When solvation is taken into account (2a and 2b), charge is always local-ized on the initial site with high probability (~0.98 at $\xi = 0$ and ~0.97 at $\xi = 6.4 \times 10^{-5}$); that is, only very small waves scatter. Probability distributions over sites for cases 1b and 2a at various time moments, when the probability at chain ends is almost zero (and does not influence charge mobility estimates), is shown in Fig. 2.

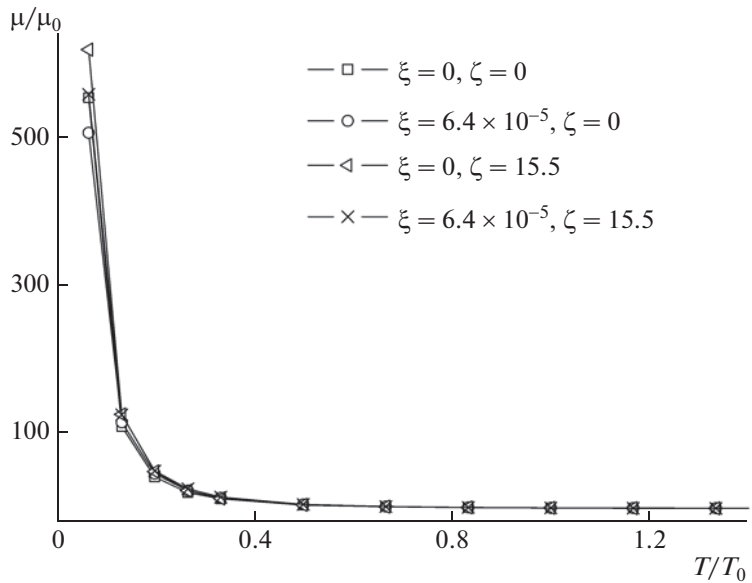

**Fig. 1.** Temperature dependence of relative mobility; $T_0 = 300$ K, $\mu_0 = \mu(T_0)$ ($\mu_0$ values are given in the table).

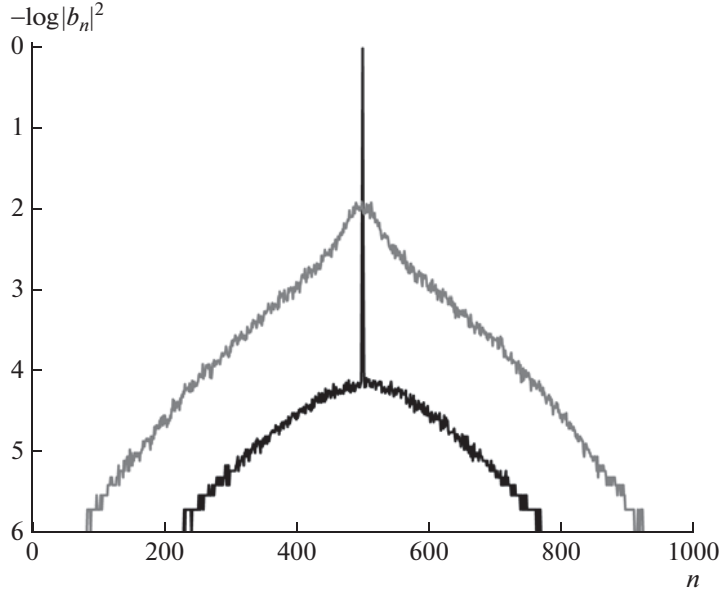

**Fig. 2.** Distributions of average probabilities over sites. Numbers of chain sites are plotted on the horizontal axis, and probabilities $|b_n|^2$ averaged over realizations, on the vertical axis. The data plotted in gray correspond to the distribution for case 1b (without solvation but with dispersion) at the $t = 1100$ calculation moment, and the data plotted in black, to the distribution for case 2a (with solvation but without dispersion) at time $t = 7000$; $T = 350$ K, averaging over 100 realizations.

## RESULTS AND DISCUSSION

Mobility estimates for a chain with solvation give larger values than earlier analytical estimates [11–13]. They are closest to the value obtained by Conwell and Basko [13], the upper mobility estimate is $\mu < 0.01$ cm$^2$/(V s). The value calculated for case 2a is $\mu = 0.052$ cm$^2$/(V s). However, note that mobility was estimated in [13] using the solvation coefficient two times smaller than we used in our calculations; it can be suggested that, at lower solvation characteristics, charge is localized more weakly, and mobility can be higher than that obtained in calculations.

Theoretical estimates [10, 22] of mobility variations in chains when dispersion changes (without taking temperature into account and in the continual approximation) give $\mu$ ($\xi = 6.4 \times 10^{-5}$)/$\mu$ ($\xi = 0$) ≈ 2.8 for DNA parameters. According to direct simulation results (see table and Fig. 1), at any temperature between 20 and 400 K, $\mu$ ($\xi = 6.4 \times 10^{-5}$)/$\mu$ ($\xi = 0$) ≈ 4.4 in a chain without solvation (the first case) and $\mu$ ($\xi = 6.4 \times 10^{-5}$)/$\mu$ ($\xi = 0$) ≈ 4.8 (the second case). These values are fairly close to theoretical estimates. In the variants under consideration, the ratios between mobilities are almost equal at various temperatures over the temperature range studied (see Fig. 1).

The contribution to calculations of the mean square displacement $\langle X^2(t) \rangle = \sum |b_n(t)|^2 n^2$ and, accordingly, diffusion coefficient and mobility is determined by the probabilities of waves $|b_n(t)|^2$ running from the initial site in the center of a chain with $n_0 = 0$ and site numbers $n$. The shapes of plots shown in Fig. 1 leads us to suggest that the rate of spreading of delocalized probability distribution fraction (which is about 1 in a chain without solvation and about 0.02 in a chain with solvation, when the main probability fraction is localized on one site, see Fig. 2) mainly depends on temperature. The absence or presence of dispersion and solvation in a chain weakly influence the temperature dependence of relative mobility $\mu/\mu_0$ ($T/T_0$).

Note that, for chains with other parameters (site frequencies $\omega$, matrix elements $\eta$, and coupling constants between the quantum and classical subsystems $\chi$), the power dependence can be different; for instance, for $\omega^2 = 0.1$, $\eta = 0.5$, $\chi = 0.632$, and friction coefficient $\gamma = 0.01$, we have $\mu/\mu_0 \approx (T/T_0)^{-1.8}$ [23].

Calculations were performed using the simplest assumption that model coefficients at all the temperatures were equal. Parameters are, of course, temperature-dependent. For instance, the interaction constant with stretching and contraction H-bond deformations $K$ strongly depends on temperature; at ~80°C (~350 K), melting of DNA occurs; H-bonds of complementary pairs then dissociate, and DNA strands untwist; that is, $K \to 0$ at $T \sim 350$ K. The dispersion constant $\xi$ depends on temperature more weakly. It can be suggested that both these parameters, $K$ and $\xi$, change more and more weakly as the temperature lowers (become constant). It can also be suggested that the effective solvation coefficient $\varsigma$ changes as water experiences the transition to the solid state. It probably decreases. Test calculations show that, at equal thermostat temperatures, mobility is higher in the chain in which the solvation coefficient $\varsigma$ is lower; that is, it can be suggested that simulation with taking the temperature dependence of molecule parameters into account would give higher mobility values than those obtained in calculations. This is the subject matter of future studies.

## CONCLUSIONS

To summarize, calculation experiments allow us to make the following suggestions. At physiological temperatures 280 K ≤ $T$ ≤ 350 K, polarons rapidly decompose in dry DNA (or have no time to form). For this reason, current carriers in dry DNA are zonal (delocalized) holes. This results in mobilities of charge carriers larger than those in "wet" DNA, in which polarons are stable and are the main current carriers. In a polynucleotide fragment, charge mobilities differ (by 2 orders of magnitude according to our estimates) depending on whether or not this DNA fragment is in solution.

The calculated mobility values for a chain with solvation are comparatively large, and solvation, although it hinders charge transfer, does not prevent it in nucleotide chains.

Calculation results are also indicative of strong influence of dispersion in a chain on the mobility value. Taking into account dispersion (DNA stacking) increases mobility by almost 5 times. To explain this, note that, for DNA, dispersion (stacking interaction) coefficients between sites and elastic coefficients ("intrasite" H-bonds) are close to each other.

Note in conclusion that, in the absence of dissipation (at $\gamma = 0$) and at $T = 0$, $\mu \to \infty$ in the Holstein model. This corresponds to dissipation-free movements of Holstein polarons in an exponentially narrow polaron zone. The presence of dissipation results in a finite mobility at $T = 0$. For real damping values $\gamma$, hole mobility is small [14, 22]. The tendency of $\mu$ to infinity as $T \to 0$ at real $\gamma$ values (Fig. 1) is related to the inapplicability of the semiclassical model used by us at $k_B T \leq \hbar\omega$ ($T \leq 10$ K) [24].

## ACKNOWLEDGMENTS

This work was financially supported by the Russian Foundation for Basic Research (project nos. 10-07-00112, 11-07-00635). The authors are grateful to the staff of the Joint Supercomputer Center of RAS for providing facilities for our calculations.